\documentclass[12pt,preprint]{aastex}

\begin{document}

\title{Quantifying the Bull's Eye Effect}

\author{Brian C. Thomas, Adrian L. Melott, Hume A. Feldman and Sergei F. Shandarin}
\affil{University of Kansas, Department of Physics and Astronomy \\
           Malott Hall, 1251 Wescoe Hall Dr, Room 1082 \\ 
           Lawrence, KS  66045-7582}

\begin{abstract}
We have used N-body simulations to develop two independent methods to quantify redshift distortions known as the Bull's Eye effect (large scale infall plus small scale virial motion).  This effect depends upon the mass density, $\Omega_0$, so measuring it can in principle give an estimate of this important cosmological parameter.  We are able to measure the effect and distinguish between its strength for high and low values of $\Omega_0$. Unlike other techniques which utilize redshift distortions, one of our methods is relatively insensitive to bias.  In one approach, we use path lengths between contour crossings of the density field.  The other is based upon percolation.  We have found both methods to be successful in quantifying the effect and distinguishing between values of $\Omega_0$.  However, only the path lengths method exhibits low sensitivity to bias.
\end{abstract}


\section{Introduction}
\label{sec-intro}
Distortions in redshift maps of galaxy positions have been known for decades.  Important work was done on linear redshift distortions during the mid 1980's \citep{kaiser87}, and more recently work has been done toward utilizing the effect to measure the linear redshift distortion parameter $\beta \approx {\Omega_0}^{0.6} / b$ \citep{hamilton98, peacock01}.  This measure can yield an estimate of the mass density, $\Omega_0$, if one has independent knowledge of the bias parameter, $b$.  In addition, \citet{hui00} have explored the probability distribution function of density in relation to redshift distortions.

It has been noted that one consequence of redshift distortions, known as the Bull's Eye effect \citep{praton94, praton97}, depends upon the value of the mass density and may be independent of $b$ \citep{melott98}.  As pointed out by \citet{bridle03}, $\Omega_0$ is a critical parameter in testing our cosmological models but is currently not known with high certainty. Measuring the strength of the Bull's Eye effect may give an alternative estimate of this important cosmological parameter without requiring knowledge of the value of the bias parameter. The only data this approach requires is redshifts and so we plan to utilize data from galaxy redshift surveys such as 2dF \citep{colless01} and Sloan Digital Sky Survey (SDSS) \citep{stoughton02}.  

Redshift distortions occur because galaxies have local (peculiar) velocities which either add to or subtract from their line of sight redshifts due to the general Hubble expansion.  The Bull's Eye effect is composed of two effects of redshift distortion: small scale stretching (known as ``Fingers of God'') and large scale compression, both along the line of sight.  Stretching is caused by the small scale random velocities of galaxies near the center of collapsed clusters.  Here, some galaxies are moving toward the observer (decreasing their redshift) and others are moving away (increasing their redshift), so the net effect is to spread the distribution out in the line of sight direction.  Compression is the result of large scale coherent motion such as infall of galaxies toward the center of a cluster or wall.  Here, galaxies on the near side of the cluster are moving away, while those on the far side are moving toward the observer, so the net effect is to compress the distribution along the line of sight. For more on both of these effects, see the thorough review of redshift distortions by \citet{hamilton98}, as well as \citet{hui00}.

\begin{figure}
\plotone{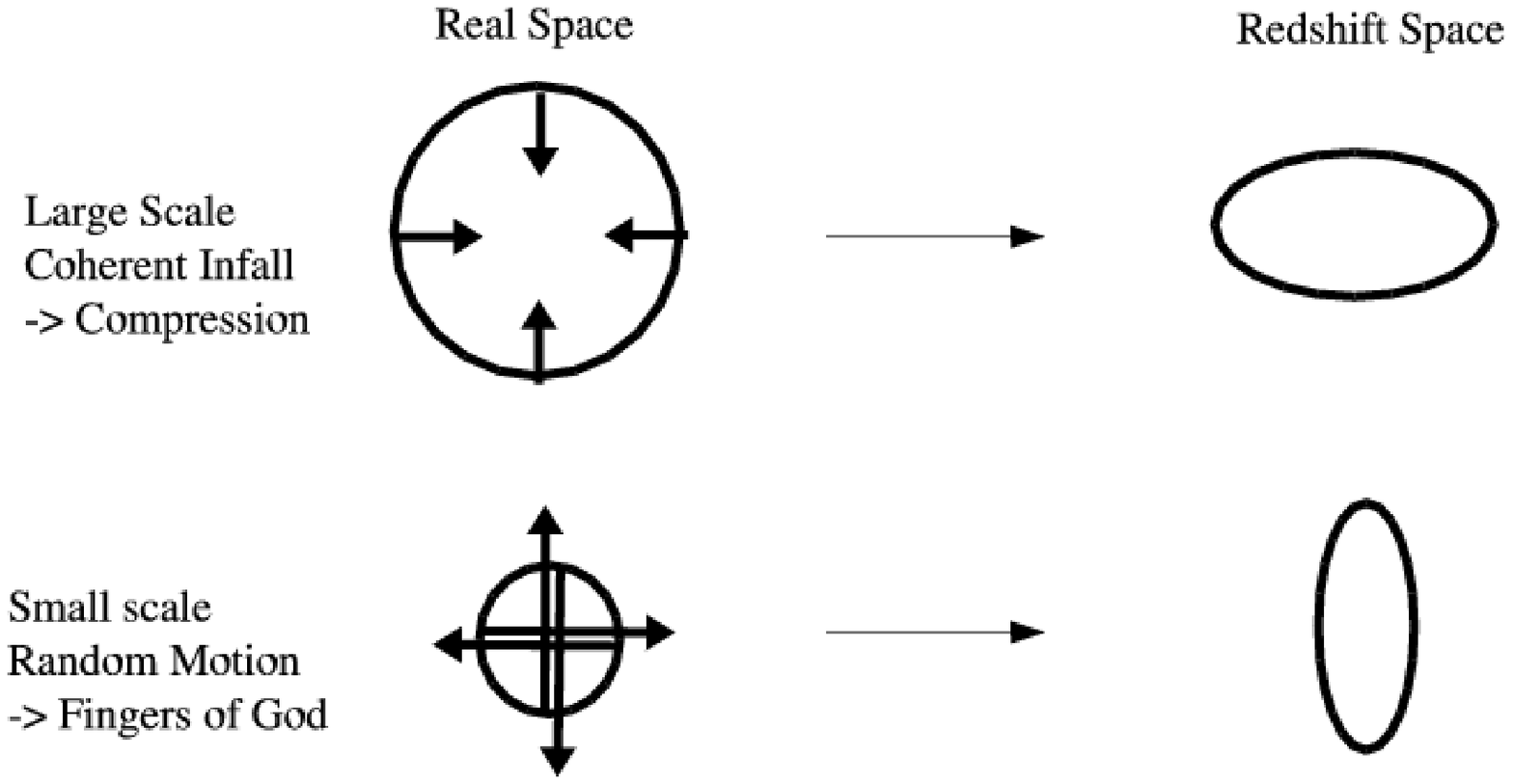}
 \caption{Compression and Fingers of God.  Large scale coherent motion leads to compression along the line of sight in redshift space and small scale random motion leads to stretching (also along the line of sight).  Here the observer is located \textit{below} the figure, looking up.  Note that transverse motion is not detected and so does not contribute to the effect.}
 \label{fig:effect}
\end{figure}

\begin{figure}
\epsscale{0.80}
\plotone{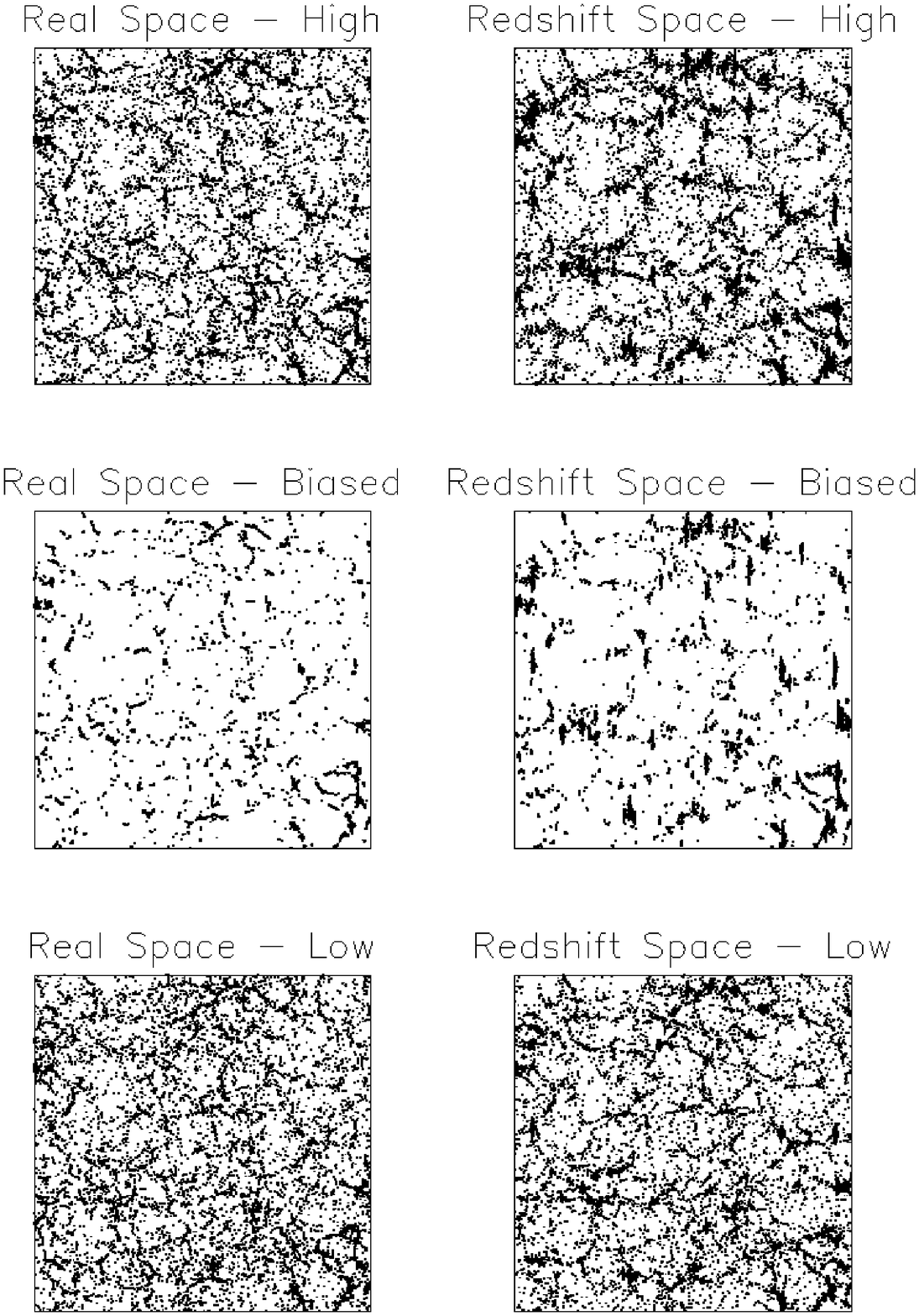}
 \caption{Effect of redshift distortions for high $\Omega_0$ (both biased and unbiased) and low $\Omega_0$ (unbiased) from our N-body simulations.  The redshift direction is vertical.  Note the presence of the compression and stretching (Fingers of God) components.  Note also the increased intensity of the effect for high vs. low $\Omega_0$.}
 \label{fig:effect_omega}
\end{figure}

The combination of these two components results in a visual effect in which the observer sees some structures pointing directly at him, and others that seem to encircle him; this is the origin of the name ``Bull's Eye.''  Figure~\ref{fig:effect} gives a schematic depiction of the two components of the effect (based upon a similar figure in \citet{hamilton98}).  Figure~\ref{fig:effect_omega} shows how the effect appears in redshift space maps of galaxy positions. These plots were made using a random subset of points from a two dimensional slice of the full three dimensional simulations (see Section~\ref{sec-methods_results} for details on the simulations).  The number of points was chosen to give approximately the same density as that of $L_*$ galaxies in the Universe.  Each plot contains approximately the same number of points.  The biased plot appears more sparse due to greater clustering.  Smoothing reduces this difference between biased and unbiased by removing points in low density areas (voids), as may be seen in Fig.~\ref{fig:contours}.

The strength of the compression part of the Bull's Eye effect depends on the value of $\Omega_0$. In fact, for our purposes here the stretching part is noise.  See Fig.~\ref{fig:effect_omega} for an example of the effect with differing values of $\Omega_0$. The connection between the strength of the effect and the value of $\Omega_0$ can be understood by using the Zel'dovich approximation \citep{zeldovich70}.  This approximation relates the final position, $\vec r$, of a particle to its initial position, $\vec q$: \begin{equation}
\vec r = a(t)[\vec{q} - D(t) \nabla_{\vec{q}} \Phi(\vec{q})]
 \label{eq:Zeldovich1}
\end{equation}
where $D(t)$ is the growing mode solution for density perturbations, $a(t)$ is the cosmic scale factor, and $\Phi$ is the primordial gravitational potential.  One can relate $D$ to $\Omega_0$ through 
\begin{equation}
f = {{d \log D} \over {d \log a}}
\end{equation}
 where $f$ can be approximated as $\Omega_0^{0.6}$ \citep{peebles80}.

The velocity $\vec v$ of an element can be found by differentiating Eq.~\ref{eq:Zeldovich1}, which yields:
\begin{equation}
\vec v = H \vec r - a(t) \dot D \nabla_{\vec q} \Phi(\vec q)
\end{equation}
This velocity can be translated into an effective distance in redshift space
\begin{equation}
d_z = v/H = r_z - fa(t)D(t) {\partial{} \over\partial{z}} \Phi(\vec q) = aq_z - (1+f)a(t)D(t) {\partial{} \over\partial{z}} \Phi(\vec q)
\end{equation}
Where $z$ denotes the redshift (line of sight) direction.  Hence the displacement exactly along the line of sight is multiplied by a factor $(1+f)$.  Since $f \approx \Omega^{0.6}$, it is apparent that the compression (along the line of sight) depends upon $\Omega_0$.

Our ultimate goal is to refine a method which can determine numerically the strength of the redshift distortion associated with a particular value of $\Omega_0$.  This method will then be applied to galaxy redshift survey data in order to determine the value of $\Omega_0$ for our universe.  Here we use N-body simulations rather than surveys in order to develop methods which probe the effect and can then be modified and applied to surveys.

Throughout the paper we will use $\Omega_0$ to denote the mass density in baryons plus cold dark matter at a redshift of zero.  The effect of a cosmological constant or ``dark energy'' component is negligible for this effect.  Also note that here ``high'' $\Omega_0$ indicates the critical value, while ``low'' $\Omega_0 \approx 0.3$.

We present two methods and their results in Section~\ref{sec-methods_results} and discuss the results in Section~\ref{sec-discussion}.

\section{Methods and Results}
\label{sec-methods_results}
We have developed two methods of quantifying the Bull's Eye effect using N-body simulations.  Each method is used to show two things: first, that we can measure the effect (for a given model); second, that we can distinguish between different values of $\Omega_0$ using that measure of the effect.  It should be emphasized that the purpose of this study is to work out methods which can probe the effect.  Therefore, we have used simulations, wherein there is less uncertainty and parameters may be easily varied in order to test our methods for a variety of possible cases.  Future work will utilize knowledge gained through this study to measure the effect using actual redshift survey data.

The simulations used here are numerical models for the gravitational dynamics of collisionless particles in an expanding background. All of the simulations used are done with a particle-mesh (PM) code with $256^3$ particles in an equal number of grid points \citep{melott86, melott88}.  The physical size of the simulation box is $512~Mpc$ on a side (for the value of $h$ given below). The simulations were normalized to an amplitude $\sigma_8=0.93$ at redshift moment $z=0$, and assume a Harrison-Zel'dovich primordial power spectrum.  The typical correlation length is about three cells ($6~Mpc$).  All models were interpreted with an assumed Hubble constant $h = H_0/100~km~s^{-1}Mpc^{-1} = {2 \over 3}$. When power spectra are parameterized in $Mpc$, the shape is dependent upon $\Gamma h$, where $\Gamma=\Omega_0 h$ is a formal shape parameter in the power spectrum formula of \citet{bbks86}.  Here, we decouple $\Gamma$ and $\Omega_0$ since it is known with reasonable certainty that the shape of the power spectrum is not too far  from $\Omega_0 h \approx 0.2$ \citep{percival02}.  Our intent in doing so is to assess the value of the Bull's Eye effect in measuring $\Omega_0$.

We do not wish to test our method \textit{only} against currently favored cosmologies. We ran a simulation with $\Omega_0 = 1$ and $\Gamma h = 0.15$; such models have been called $\tau$CDM in the past. This is our ``high $\Omega_0$'' model.  We also ran a simulation with $\Omega_0 = 0.34$, $\Gamma h = 0.15$, which is most consistent with a variety of findings at this time. This is our ``low $\Omega_0$'' model.  There are, in addition, a variety of alternative models with the cosmological constant, $\Lambda$, all of which have very small and totally linear effects on large-scale velocities, especially for $\Omega_0 \ge 0.3$ \citep{hamilton01}.  Therefore, we omit this in favor of wider exploration of parameter shifts with large effects.

For each simulation, redshift space versions were constructed, where the redshift distortion is included along one of the three orthogonal directions.  In addition, we have versions where bias is included.  There are many ways to produce a biased numerical simulation.  In this case, particles were tagged if, in the initial conditions, they lay in regions of density enhanced 1$\sigma$ above the mean on a scale of R = 1 $Mpc$ (Cloud in Cell smoothing \citep{hockney81}).  Using only these particles to define the final state defines the biased density field.  The bias factor $b$ = 1.8 for this procedure, where $b$ is defined as the ratio of $\sigma_8$ in the biased density field to that in the underlying simulation from which it was constructed.  While the value of the bias parameter in our universe is not known, our $b$ is comfortably larger than any indicated by current data so that we can put a safe upper limit on the extent to which this is likely to be a problem for our method.

For our analysis, we used three realizations of each simulation (differing only in their initial random number seed).  Each of these realizations has three redshift space versions (with the redshift distortion along one of the three orthogonal directions) and one real space version.  The analysis is performed using two dimensional slices of the three dimensional simulation density field, in Cartesian coordinates.  A slice is one grid cell thick and is 256 cells on a side.  From each redshift space simulation box we take 256 slices containing the redshift distortion.  Specifically, for redshift along the x and y directions, we take slices at constant z (x-y planes); for redshift along the z direction we take slices at constant x (y-z planes).  For the real space boxes we take 256 slices along each direction.  Therefore, we have (3 realizations)x(3 redshift directions)x(256) redshift slices and (3 realizations)x(3 directions)x(256) real space slices available.  

From this set of slices we compile an ensemble to use for the analysis.  This ensemble is obtained by keeping only every fourth slice from each simulation box in order to reduce correlation between slices due to proximity.  This yields an ensemble of 3x3x(256/4)=576 slices (for a given model), which, when taken together, comprises slightly less volume than that of the 2dF survey \citep{colless01}.

Slices of the density field have been used since the surveys we will be applying our methods to (such as 2dF and SDSS) include thin slice regions.  This also provides a large number of realizations to our ensemble.  Cartesian coordinates have been used in order to facilitate development of a method which can then be adapted to apply to polar data such as that from redshift surveys.  A Cartesian approach can, in principle, be used directly for a deep enough survey where a plane parallel approximation may be used and may lead to a statistic which is usable in polar coordinates. Again we emphasize that our purpose here is to work out a method in a simpler situation which may then lead to a method suitable for use with real data.

Before any analysis is done, the density slices are smoothed by convolution with a Gaussian $\sim e^{-r^2/2\lambda^2}$.  This smoothing serves two purposes.  First, it ensures that the density does not depend upon grid size in the simulations, since the grid is smaller than the typical smoothing length. Second, smoothing helps to compensate for greater noise in real data as compared to simulations, due to a lower density of galaxies in real data.  Typical galaxy separation in the universe is $3~Mpc/h$ ($4.5~Mpc$ for $h = 2/3$) while in our simulations the separation is $2~Mpc$.  The vast majority of galaxies are clustered, as evidenced by the fact that most galaxies have a nearest neighbor closer than the mean separation \citep{ryden84,linder96}.  Therefore, smoothing emulates the real (clustered) galaxy distribution better. 
 
The smoothing length,~$\lambda$, is set to be a fraction of the correlation length, $r_0$, of the slice.  The correlation length is used to set the smoothing length because it is stable and easy to measure.  All the data presented here are for a fraction of $1/2$.  We note that the presence of bias changes the correlation length and so will imply an effect on the smoothing length.  Therefore, investigation of variability of the smoothing length is accounted for by examining the biased simulations.

\subsection{Path length statistics}
\label{sec-methods_results-paths}
\begin{figure}
\epsscale{0.8}
\plotone{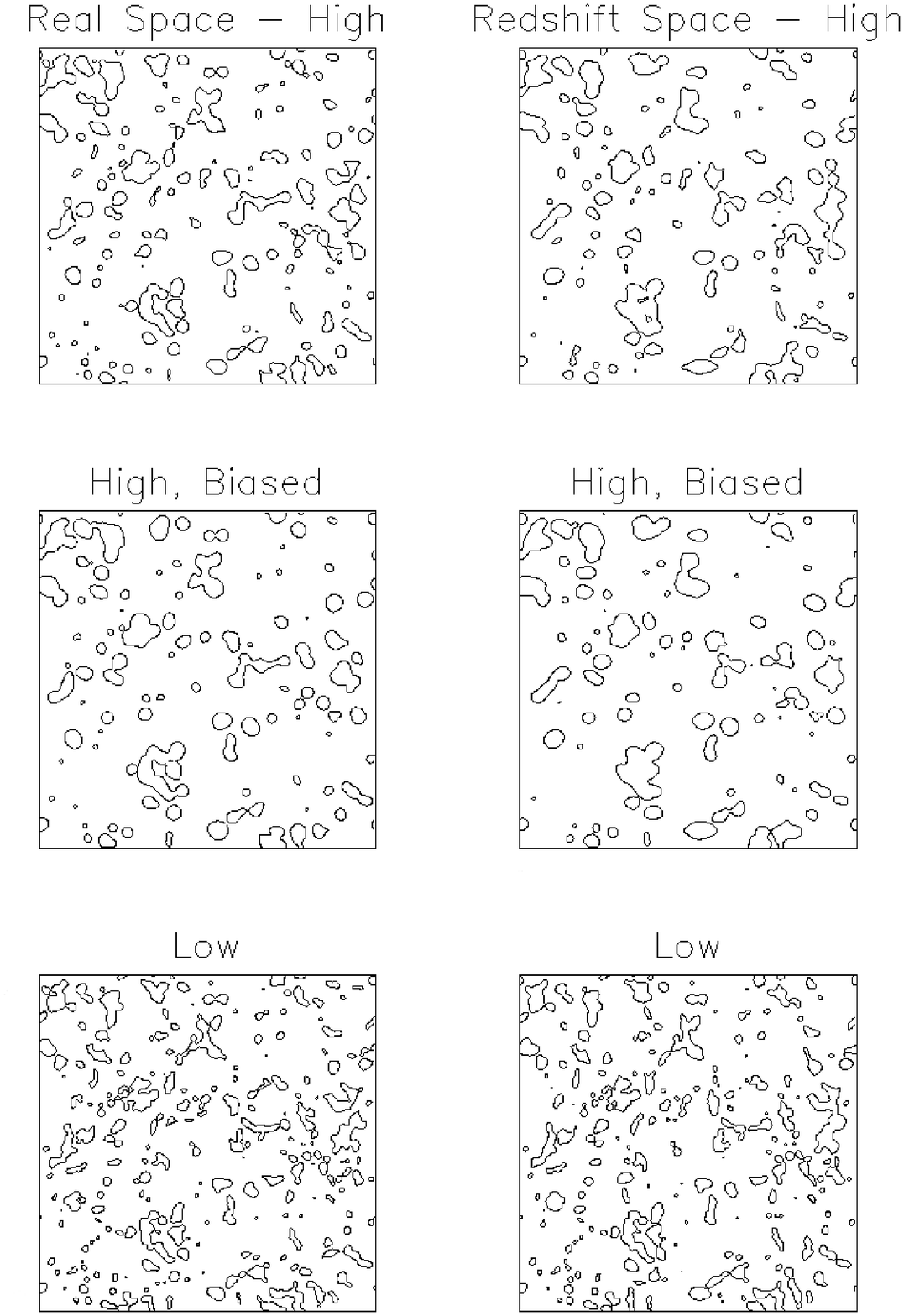}
 \caption{Contour plots of a typical 2D slice of the density field at filling factor 0.15.  Left column is real space, right column is redshift space, where the redshift direction is \textit{vertical}.  Top row is high $\Omega_0$, middle is biased high $\Omega_0$, bottom is low $\Omega_0$.}
 \label{fig:contours}
\end{figure}

Our first approach involves using contour crossings of the 2D slices of the 3D density field.  Contour plots of a typical slice are shown in Fig.~\ref{fig:contours}. Real and redshift space (with redshift in the vertical direction) are shown for comparison, with high $\Omega_0$, biased high $\Omega_0$ and low $\Omega_0$. In these plots the contours are made for a fixed value of the filling factor, $\phi$, which is defined as the fraction of the total number of sites which have densities above a given threshold.  That is, the actual density level used varies between slices and $\Omega_0$ values for a specific value of $\phi$. Using filling factor instead of density threshold reduces our sensitivity to bias \citep{melott88}.  Biasing increases the amplitude of fluctuations.  Therefore, using a fixed fraction of the density above a threshold tends to normalize biased to unbiased density fields (as long as biased density increases monotonically with unbiased density).  This can be seen in Fig.~\ref{fig:contours} by comparing the biased and unbiased contours which appear similar here.  Note, however, that biasing also increases the correlation length and therefore (since we use a fraction of $r_0$ as the smoothing length) the biased density field is effectively smoothed at a larger scale than the unbiased.  This is also apparent in Fig.~\ref{fig:contours}.

\begin{figure}
\epsscale{1}
\plotone{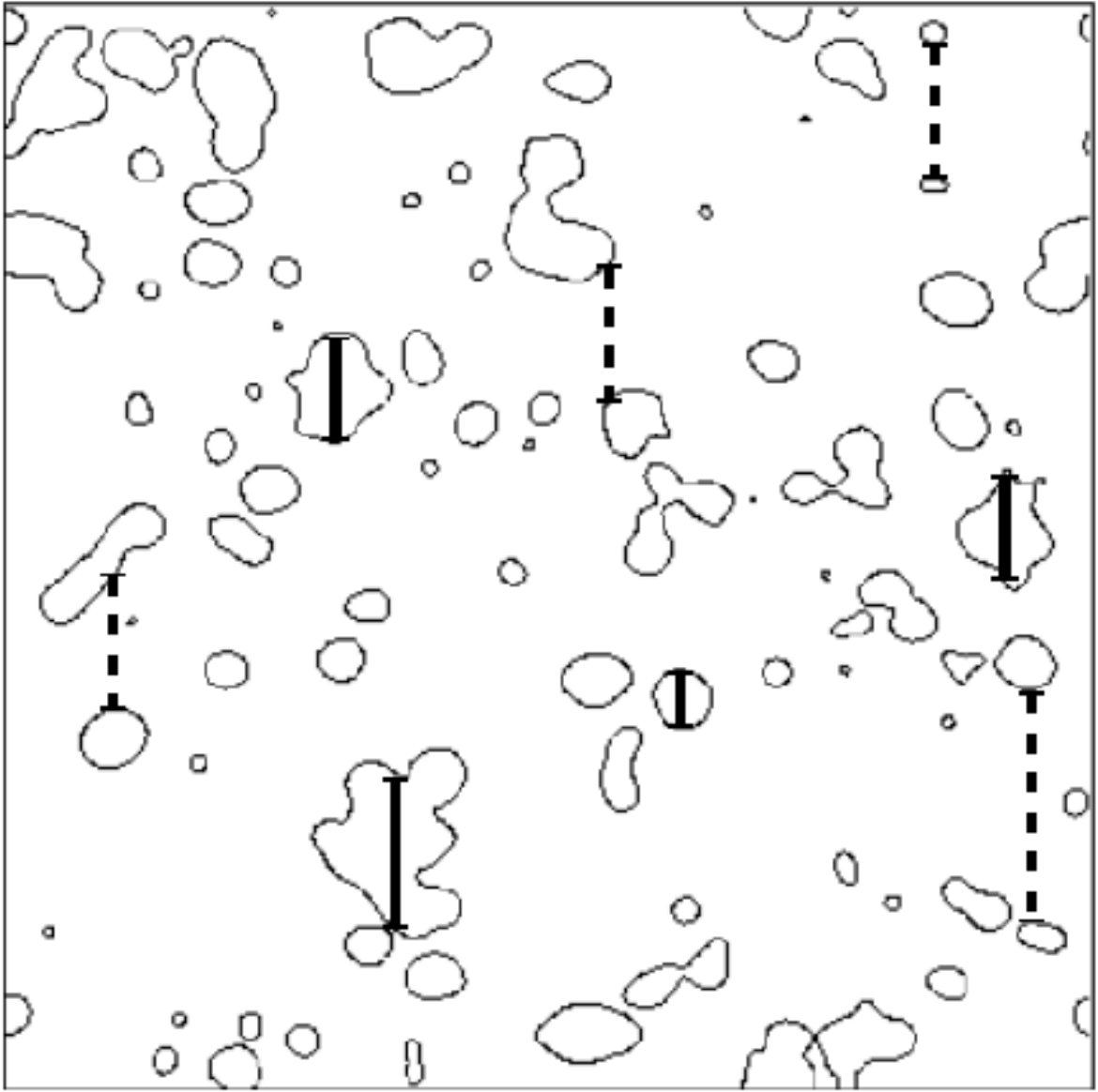}
 \caption{Examples of paths marked between particular contour crossings.  Solid lines are ``structure paths,'' marked between an up-crossing (from below the threshold to above) and a down-crossing (from above the threshold to below).  These paths measure the size of structures.  Dashed lines are ``void paths,'' marked between a down-crossing and an up-crossing, measuring the size of voids.}
 \label{fig:paths}
\end{figure}

The analysis is done by looking at the contours at a given filling factor and marking out \textit{paths} (in a certain direction; ie., x or y) between contour crossings \citep{ryden88, ryden89}.  Various different paths can be marked, in particular those measuring structures and voids. Structure paths are marked between a contour crossing where the density increases from below the threshold to above (an \textit{up-crossing}) and a crossing where the density decreases from above the threshold to below (a \textit{down-crossing}). Hence, a structure path is an ``up-to-down'' path.  Conversely, void paths are marked between a down-crossing and an up-crossing; they are ``down-to-up'' paths.  See Fig.~\ref{fig:paths} for examples of both types of paths.

Our analysis of these paths proceeds as follows. We start with individual path lengths in a given slice (at a given filling factor), $l^{T}_{i,j}$ and $l^{L}_{i,j}$, where superscripts denote path direction ($T$~=~transverse, $L$~=~line of sight), subscript $i$ denotes a path and subscript $j$ denotes a slice.  We then calculate the mean of these path lengths in a single slice,
\begin{equation}
\bar l^{T}_{j} = {1 \over n^{T}_j} \sum_i l^{T}_{i,j} \;\;\;,\;\;\; \bar l^{L}_{j} = {1 \over n^{L}_j} \sum_i l^{L}_{i,j}\;\;,
\end{equation}
where $n^{T}_j$ and $n^{L}_j$ are the number of paths in each direction for slice $j$ (ranges from about 70 at $\phi=0.01$ to about 500 at $\phi=0.5$).  In addition to the mean, we have investigated the dispersion and rms of path lengths as possible probes of the effect.  The mean was found to give the strongest signal.

Once the mean of path lengths in a slice is found, we calculate for each slice the ratio of mean path lengths in the line of sight to transverse directions,
\begin{equation}
r_j = {\bar l^{L}_{j} \over \bar l^{T}_{j}}.
\end{equation}
We then find the median value of this ratio for the ensemble of 576 slices previously described.  (The median is used because the distribution of ratios is non-Gaussian.) For use with this method, we record only paths from every fourth column (in both directions) in order to reduce correlation between paths within a slice. For most slices this distance (four cells) is similar to or slightly larger than the correlation length, which typically is between three and four cells.

Once the median of the ratios of mean path lengths has been calculated, it is plotted against filling factor, $\phi$.  This type of plot is shown in Fig.~\ref{fig:path-stats-4plot} for real and redshift space, for both structure and void paths. In these plots, the error bars (denoted by thin lines) correspond to 68\% confidence intervals of the median.  
This is the confidence in the median for a measurement including a single slice.  For the volume (all slices taken together), the size of the errors are reduced by a factor of $\sqrt{n}$, where $n$ is the number of slices.
For the void paths, empty columns in the simulation box present a problem, since they are essentially paths of indeterminate length (due to periodic boundary conditions). We have, therefore, chosen to assign empty columns a length of twice the box size.  This gives us paths which are larger than the box (as they must be), but which are not unrealistically large.  Inclusion of empty columns in this way has very little effect on the results.

A general characteristic of the redshift space plots in Fig.~\ref{fig:path-stats-4plot} (for both void and structure paths) is that at low filling factor the ratio is greater than one and at higher filling factors the ratio is less than one.  (The $\phi$ value where this transition occurs varies for the different models.)  
For the void paths, a ratio greater than one indicates that the mean of void size is greater in the line of sight direction (as was also found by \citet{ryden96} and \citet{schmidt01}) as compared with the transverse direction. This is the result of the distance between widely separated structures becoming larger due to large scale coherent motion, which leads to an increase in the size of large voids (which are the majority at small filling factors).  This, in turn, results in the shrinking or disappearance of small voids from their compression by larger voids \citep{melott83}.  At higher filling factors, where the ratio dips below one, the majority of voids are small and hence the shrinking of small voids results in a shorter mean path length in the line of sight direction compared with that in the transverse direction.

For the structure paths, a ratio greater than one indicates that the mean of structure size is greater in the line of sight direction as compared with the transverse direction. This is caused by the Fingers of God effect due to small scale random motion, which lengthens the structures in the line of sight. At higher filling factors, where the ratio dips below one, the structures are larger and the compression effect dominates (since it is a large scale effect), compared to low filling factor where the structures are smaller and hence the small scale effect (Fingers of God) dominates.

We note that comparison of the redshift and real space plots in Fig.~\ref{fig:path-stats-4plot} clearly shows that this method detects the presence of the Bull's Eye effect in redshift space.  Our expectation of no detection of the effect in real space is confirmed since the ratios are at or near 1 (in contrast to redshift space ratios) for all models.

In this study, we wish to probe the large scale, coherent motion (which leads to compression of structures) since it is primarily this effect which, by physical reasoning, should be dependent upon the value of $\Omega_0$, as described in Section~\ref{sec-intro}. Conversely, small scale, random motion (which produces Fingers of God) is noise for our purposes.  Therefore, further discussion of the analysis of path lengths will focus on the void paths since, as noted above, these paths are more sensitive to compression while structure paths are more strongly affected by Fingers of God.

Before we proceed we must comment on the peak which appears in the upper left panel of Fig.~\ref{fig:path-stats-4plot} (void paths in redshift space).  The turn-over which occurs around filling factor 0.03 appears to be unphysical. We believe this to be so for two reasons:  1) Extremely low filling factor values correspond to a small percentage of the density above the threshold, yielding few structures to use as references for voids, leading to a small number of void-paths. 2) The scarcity of structures, as well as their small size, means that the majority of void paths present are large compared to the simulation box size (greater than half the box in length).  For instance, at $\phi=0.01$, just to the left of the peak, a typical slice has about 70 paths, 90\% of which are greater than half the box.  At the peak ($\phi=0.03$), a typical slice has about 100 paths, 60\% of which are large. At $\phi=0.05$, just to the right of the peak, there are about 120 paths, 50\% of which are large.  Moving further to the right, at $\phi=0.15$ a typical slice has about 280 paths, 10\% of which are large.

It is apparent, then, that at small filling factors (less than about 0.03) any difference in path length between the line of sight and transverse directions is too small of a fraction of the total length to be significant.  This conclusion is strengthened by the fact that there is no turn-over at low filling factors for the structure paths, where small $\phi$ yields few paths, none of which are long (since structures are few and small).  We conclude, therefore, that the turn over in the upper left panel of Fig.~\ref{fig:path-stats-4plot} is unrealistic and our analysis will only be applied to filling factors greater than 0.03.

In addition to detecting its presence, we wish to use the strength of the Bull's Eye effect to estimate the value of the mass density.  Therefore, we must determine whether the method at least distinguishes between widely separated values of $\Omega_0$.  Further, we look for the method to have a low sensitivity to bias.  For a certain range of filling factor in the redshift space, void paths plot in Fig.~\ref{fig:path-stats-4plot} (upper left panel), it is apparent that high and low $\Omega_0$ can be distinguished.  In addition, in the range $\phi \approx 0.1$ to $\phi \approx 0.15$ biased and unbiased high $\Omega_0$ are nearly identical.  Therefore, in the filling factor range $\phi \approx 0.05$ to $\phi \approx 0.2$ the method both distinguishes between widely separated values of $\Omega_0$ and is reasonably insensitive to bias.  We have tested the method using a variety of power spectra and the quoted $\phi$ range is consistently best.

In order to more precisely determine how well the method distinguishes between $\Omega_0$ values, as well as how sensitive it is to bias, we have applied a Wilcoxon rank-sum test to the ensembles of ratios of mean path length.
 This test was used because it determines whether one set of data is systematically \emph{not greater} in magnitude than another (for details, see \citet{bluman01}~and~\citet{lehmann98}). 
In this case, we expect the ratios of mean path length to be systematically larger for high $\Omega_0$ (both biased and unbiased) compared to low $\Omega_0$.  As a test for sensitivity to bias, we compare between biased and unbiased high $\Omega_0$.

We have applied this test to void paths in redshift space, at a given filling factor. We use the same ensembles of ratios used to produce the median plot in the upper left panel of Fig.~\ref{fig:path-stats-4plot}.  This gives us 576 ratios (for a given model), produced from every fourth slice of our simulation boxes.  In choosing a value of the filling factor at which to apply the test, we must balance the desire for a strong signal with the need to be insensitive to bias.  Therefore, we take $\phi \approx 0.1$ as a good choice since the signal is relatively large here and both the biased and unbiased high $\Omega_0$ give similar results.  This is true for both void and structure paths, indicating reliability of this choice of filling factor value. 

In Table~\ref{tbl-paths_wil} we present results of comparing between values of $\Omega_0$ for void paths in redshift space at filling factor 0.1 using the Wilcoxon rank-sum test.  That is, we test to see how well the method distinguishes between models.
 For each comparison we give the probability that the ratios of mean path length for model A are \emph{not} greater than those for model B.  Therefore, small P indicates that the ratios for A \emph{are} greater than those for B.
We see, then, that the ratios for high $\Omega_0$ (both biased and unbiased) are greater than for low $\Omega_0$ with confidence approaching 100\%.  We also note that sensitivity to bias is not great, as there is only 88\% confidence that the ratios for unbiased high $\Omega_0$ are greater than those for biased high $\Omega_0$.

As an additional check on these results, we have applied a Kolmogorov-Smirnov (K-S) test to the same data.  The K-S test determines whether two sets of data are drawn from the same distribution (for details, see \citet{lehmann98}).  In this case, we find that low $\Omega_0$ \emph{cannot} be distinguished from both biased and unbiased high $\Omega_0$ with less than $10^{-5}$~\% confidence. That is, low \emph{can} be distinguished from high (both biased and unbiased) with confidence approaching 100\%.  Also, biased and unbiased high $\Omega_0$ \emph{cannot} be distinguished with about 33\% confidence.  That is, they \emph{can} be distinguished with only about 67\% confidence.  These results are consistent with those described above for the Wilcoxon rank-sum test.

We have shown, therefore, that our path lengths method both detects the presence of the Bull's Eye effect (as seen in Fig.~\ref{fig:path-stats-4plot}) and distinguishes between values of $\Omega_0$ with high confidence. In addition, it is reasonably insensitive to bias.

\subsection{Percolation}
A second method we have employed utilizes a percolation routine developed by \citet{dominik92}. Given a density threshold the routine denotes sites above the threshold as occupied and then checks to see whether each occupied site is part of a cluster.  The criterion for a site becoming part of a cluster is that it be adjacent to another occupied site. Diagonal sites may also be connected if the mean density of four closest sites is above the threshold.  The routine then finds the largest cluster and checks whether it spans the density slice in each (orthogonal) direction.  The threshold is systematically varied to determine the highest density level at which the largest cluster spans in each direction. For our analysis, we then convert this threshold to a filling factor value.  As discussed in Section~\ref{sec-methods_results-paths}, using filling factor reduces sensitivity to bias, which is true for this analysis as well.  For a real space density field, the filling factors at which the largest cluster spans the slice in each orthogonal direction should be close together.  This is due to the structures being isotropic.  On the other hand, for a redshift space field the filling factors should be farther apart.  This is because isotropy is destroyed by redshift distortions.  Largest cluster spanning in redshift space occurs at a higher filling factor in the line of sight direction than in the transverse direction. The difference in filling factor depends on the strength of the redshift distortion and therefore on $\Omega_0$.  

Our analysis utilizes the difference in filling factor at which the largest cluster spans the density slice in the line of sight and transverse directions.  Specifically, we take $\Delta\phi = \phi_L - \phi_T$, where $L$~=~line of sight and $T$~=~transverse.  From the ensemble of slices, we assemble lists of $\Delta\phi$ corresponding to different models in both real space and redshift space.  We wish to make two comparisons between these lists of filling factor differences.  First, to show a detection of the effect, we compare between $\Delta\phi$ in real space and in redshift space for a given model.  Second, to determine how well the method distinguishes between values of $\Omega_0$ (and how sensitive it is to bias), we compare $\Delta\phi$ in redshift space for different models.

In order to quantify these comparisons we have applied a Wilcoxon rank-sum test to the lists of filling factor differences.  As in Section~\ref{sec-methods_results-paths}, this test was used because it determines whether one set of data is systematically \emph{not greater} in magnitude than another. In this case, we expect the filling factor differences to be larger for redshift space compared to real space and for high $\Omega_0$ compared to low $\Omega_0$. 
This method is applied to the same ensemble of 576 slices as the previous method.

In Table~\ref{tbl-perc_real-red_wil} we present results comparing filling factor differences between real space and redshift space using the Wilcoxon rank-sum test.  That is, we test to see how well the method detects the Bull's Eye effect.
We perform the test for each model and give probabilities that the filling factor differences for redshift space are \emph{not} greater than those for real space.  Therefore, small P indicates that the $\Delta\phi$ for redshift space \emph{are} greater than those for real space.
For each model we have confidence approaching 100\% that the $\Delta\phi$ for redshift space are systematically greater than for real space.  
From these results, it is apparent that we have a strong detection of the effect by this method.  The test indicates that redshift and real space can be strongly distinguished.  Sensitivity to bias in this case appears to be low since the results for biased and unbiased high $\Omega_0$ are the same.

In Table~\ref{tbl-perc_omega_wil} we present results comparing models in redshift space using the Wilcoxon rank-sum test. That is, we test to see how well the method distinguishes \emph{between} models.
For each comparison we give the probability that the filling factor differences for model A are \emph{not} greater than those for model B.  Therefore, small P indicates that the $\Delta\phi$ for A \emph{are} greater than those for B.
The $\Delta\phi$ for high $\Omega_0$ are greater than low $\Omega_0$ with 99.99997\% confidence.  
Conversely, the $\Delta\phi$ for biased high $\Omega_0$ are \emph{not} greater than low $\Omega_0$ with 99.95\% confidence.  This result is contrary to our expectation that high $\Omega_0$ will have $\Delta\phi$ greater than those for low $\Omega_0$, regardless of bias. Similarly, the $\Delta\phi$ for unbiased high $\Omega_0$ are greater than those for biased high $\Omega_0$ with confidence approaching 100\%.  Hence, this method exhibits a high sensitivity to bias.

As an additional check on these results, we have applied a K-S test to the same data. In this case, the test shows that the three models \emph{cannot} be distinguished from each other with less than $10^{-5}$~\% confidence.  That is, all the models \emph{can} be distinguished from each other with confidence approaching 100\%.  The same is true for the comparison between real space and redshift space (for a given model).  The K-S test simply determines whether two data sets can be distinguished.  Hence, the Wilcoxon rank-sum test is more informative for our purpose since we wish to determine the relative magnitude of the filling factor differences for various models.

We have shown that our percolation method both detects the presence of the Bull's Eye effect and distinguishes between high and low $\Omega_0$ with high confidence.  This method is, however, strongly sensitive to bias, as shown by the comparison between biased high $\Omega_0$ and unbiased low $\Omega_0$.  Sensitivity to bias is further seen in the strong distinction between biased and unbiased high $\Omega_0$.  It is therefore apparent that while this method may distinguish between values of $\Omega_0$, it can be deceived by the presence of bias.

\section{Discussion}
\label{sec-discussion}
We have found two distinct methods which quantitatively show the effect of redshift distortions known as the Bull's Eye effect.  These methods are also capable of distinguishing between the effect for high and low values of $\Omega_0$ with high confidence, and the path lengths method is reasonably insensitive to bias. 

Our results may be compared to measurements of redshift distortions using the galaxy-galaxy correlation function.  While the correlation function does measure the compression of the galaxy distribution along the line of sight in redshift space, we have employed different methods which measure the same effect.  The correlation function is sensitive to the combination $\beta \approx {\Omega_0}^{0.6} / b$ \citep{peacock01, hoyle01}, while we expect our approach to be sensitive to $\Omega_0$ \emph{independent} of $b$ (as shown in Section~\ref{sec-intro}).  This expectation is borne out for our path lengths method, where we see little sensitivity to bias.  The difference here could be that the correlation function is only sensitive to the fluctuation amplitude (not phases) and hence is affected by bias (which affects the amplitude), while our path length method apparently detects more than just amplitude.

\begin{figure}
\plotone{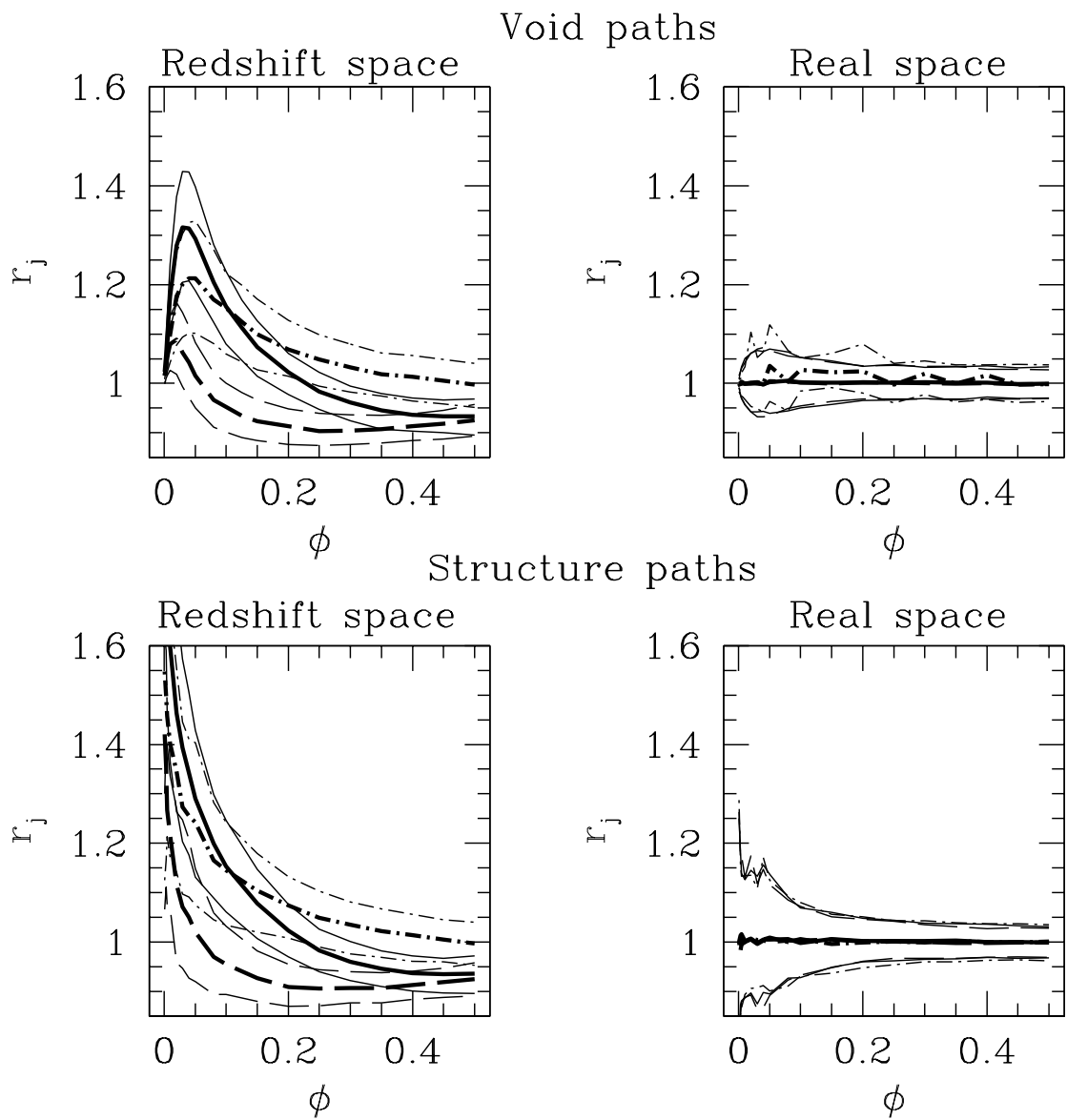}
 \caption{Plots of the median of ratios of mean path lengths.  Top row shows data for void paths, bottom row for structure paths.  Left column is data for redshift space, right for real space. In each plot: solid -- high $\Omega_0$, unbiased; dash-dot -- high $\Omega_0$, biased; long dash -- low $\Omega_0$, unbiased. Error bars (thin lines) correspond to 68\% confidence levels of the median}
 \label{fig:path-stats-4plot}
\end{figure}

The results of the path lengths method applied to void paths indicates that large voids are elongated in the line of sight direction in redshift space (see Section~\ref{sec-methods_results-paths} and Fig.~\ref{fig:path-stats-4plot}).  This effect was also noted by \citet{ryden96} and \citet{schmidt01}.  However, in both of these papers the effect was found to be relatively minor.  In comparison, our results show a significant detection of the elongation of large voids along the line of sight direction.

It should be noted that for the percolation method, the filling factors at which largest cluster spanning occurs are typically around 0.45.  We have seen with the path lengths method that biased and unbiased high $\Omega_0$ diverge at these high filling factors.  Hence, it may be reasonable to expect that the percolation method will be sensitive to bias as we have seen.

Overall, we have shown that it is possible to quantify the Bull's Eye effect in N-body simulations.  This indicates that we should be able to do the same in redshift surveys and thereby estimate the value of the mass density,  $\Omega_0$.  In addition, since the path lengths method is relatively insensitive to bias, we can therefore give an estimate of $\Omega_0$ without knowledge of the bias parameter, $b$, unlike methods which use redshift distortions to measure the distortion parameter, $\beta$.

Future work will concentrate on adapting our methods to apply to redshift survey data and then using these adapted methods to estimate $\Omega_0$ for our universe.

\acknowledgments
Thanks to E. Praton for helpful comments on the manuscript. We are grateful for supercomputing support from the National Center for Supercomputing Applications.

This work was supported by NSF grant AST-0070702 and the Madison and Lila Self Graduate Fellowship at the University of Kansas.

\clearpage

\clearpage
\begin{deluxetable}{ccc}
\tablecaption{Results of Wilcoxon rank-sum test comparing ratios of mean path lengths between models in redshift space for void paths, at filling factor 0.1. \label{tbl-paths_wil}}
\tablewidth{0pt}
\tablehead{
\colhead{$\Omega_0$} & \colhead{P} }
\startdata
High vs. Low &vs \\
High (biased) vs. Low &vs \\
High vs. High (biased) &0.1118 \\
\enddata
\tablecomments{``P'' is the probability that the ratios of mean path length for one model are \emph{not} greater than those for another.  ``vs'' indicates the probability is \emph{vanishingly small} (less than $10^{-7}$).}
\end{deluxetable}




\clearpage
\begin{deluxetable}{ccc}
\tablecaption{Results of Wilcoxon rank-sum test comparing percolation filling factor differences between real and redshift space.\label{tbl-perc_real-red_wil}}
\tablewidth{0pt}
\tablehead{
\colhead{$\Omega_0$} & \colhead{P} }
\startdata
High &vs  \\
Low &vs \\
High (biased) &vs \\
\enddata
\tablecomments{``P'' is the probability that the redshift space filling factor differences are \emph{not} greater than those for real space.  ``vs'' indicates the probability is  \emph{vanishingly small} (less than $10^{-7}$).}
\end{deluxetable}

\clearpage
\begin{deluxetable}{ccc}
\tablecaption{Results of Wilcoxon rank-sum test comparing percolation filling factor differences between models in redshift space.\label{tbl-perc_omega_wil}}
\tablewidth{0pt}
\tablehead{
\colhead{$\Omega_0$} & \colhead{P} }
\startdata
High vs. Low &$2.63\times10^{-7}$ \\
High (biased) vs. Low &0.9995 \\
High vs. High (biased) &vs \\
\enddata
\tablecomments{``P'' is the probability that the filling factor differences for one model are \emph{not} greater than those for another.  ``vs'' indicates the probability is  \emph{vanishingly small} (less than $10^{-7}$).}
\end{deluxetable}

\end{document}